# An Energy-Efficient Mixed-Signal Parallel Multiply-Accumulate (MAC) Engine Based on Stochastic Computing

Xinyue Zhang, Jiahao Song, Yuan Wang*, Yawen Zhang, Zuodong Zhang,
Runsheng Wang* and Ru Huang
Institute of Microelectronics and Key Laboratory of Microelectronics Devices and Circuits (MoE)
Peking University, Beijing 100871, P.R. China
* Email: wangyuan@pku.edu.cn, r.wang@pku.edu.cn

**Abstract**

Convolutional neural networks (CNN) have achieved excellent performance on various tasks, but deploying CNN to edge is constrained by the high energy consumption of convolution operation. Stochastic computing (SC) is an attractive paradigm which performs arithmetic operations with simple logic gates and low hardware cost. This paper presents an energy-efficient mixed-signal multiply-accumulate (MAC) engine based on SC. A parallel architecture is adopted in this work to solve the latency problem of SC. The simulation results show that the overall energy consumption of our design is 5.03pJ per 26-input MAC operation under 28nm CMOS technology.

**1. Introduction**

Excellent performances are achieved on various tasks such as image recognition and natural language processing through methods based on convolutional neural network (CNN), but at the cost of computational complexity and high power consumption [1]. Due to strict requirements for hardware resources and power consumption on edge applications, a new computing paradigm is in urgent need.

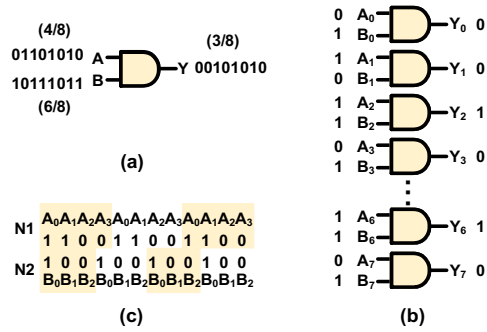

Stochastic computing (SC) is a promising candidate which can simplify the implementations of arithmetic operations and reduce power consumption [2]. Mathematical operations among stochastic numbers can be realized with simple logic gates. For instance, multiplication can be completed with a single AND gate (Fig. 1(a)). Thus SC is successfully applied to computation-intensive applications like digital signal processing [3], artificial neural networks [4] and decoding of modern error-correcting codes [5]. Convolution is an essential but energy-intensive operation in CNN which demands plenty of multiplications and additions. Since SC can simplify arithmetic operations into simple logic gates, a SC-based CNN is of great attraction.

On the other hand, high latency caused by long bit stream becomes the bottleneck of SC in real-time applications, which could be solved by parallel structure (Fig. 1(b)). Therefore, a deterministic coding method of SC is applied in this work to realize parallel structure and to remove inherent randomness. However, increasing degree of parallelism will lead to high power consumption when using a high-fan-in digital adder tree [6]. To address the problem, we propose a mixed-signal multiply-accumulate (MAC) engine to reduce power consumption.

Fig. 1. Stochastic computing. (a) Multiplication of serial structure; (b) Multiplication of parallel structure; (c) Deterministic code

The rest of the paper is organized as follow: Section 2 introduces the background and motivation of this work; Section 3 describes the structure of the proposed MAC engine; Section 4 shows the simulation results; Finally, conclusions are drawn in section 5.

**2. Background and Motivation**
**2.1 Background**

SC technique processes data in form of bit streams, where the probability of observing 1 in the stream is treated as the value of the stochastic number. For example, bit stream A=01101010 contains four 1s and four 0s, which means the value of A is 4/8 (Fig. 1(a)). Multiplication in SC can be realized with a single AND gate. Similarly, addition can be implemented with only a MUX gate. Therefore, it is promising for some computing-intensive applications to reduce logic complexity with SC technique.

However, conversion between stochastic bit streams and binary numbers do consume high power consumption and have a latency problem [7]. Besides, the nature of randomness in SC could cause some inherent errors in computation. Thus deterministic code [8] of SC has been proposed to produce completely accurate result. Fig. 1(c) shows two stochastic numbers N1 and N2, where the effective bit length of N1 is 4 and N2 is 3. Both the lengths of two numbers are extended to 12, the least common multiple of 3 and 4. We can observe that every bit of N1 sees every bit of N2, so that the calculation result will be completely accurate. Moreover, binary number can be converted into deterministic code with simple decoder in parallel [9], which highly improves the energy efficiency and reduces the computing latency.

**2.2 Motivation**

MAC is an essential but energy-intensive part in CNN, while SC technique can simplify complex computation units to simple logic gates. Therefore, a SC-based MAC engine is a promising implementation.

However, long bit stream brings an increase on computation time, and the bit stream length shows exponential relationship with the precision of the stochastic number. As shown in Fig. 1(b), this problem can be solved by increasing degree of parallelism [2]. As the degree of parallelism increases, high-fan-in addition becomes the main bottleneck when summing up all of the outputs of AND gates in SC-based convolution networks. Thus in this work, an analog addition structure is brought up to increase the energy efficiency.

**3. Circuit Description**

Fig. 2 shows the operation diagram of convolution calculation with proposed mixed-signal MAC engine, which is composed of three parts: binary-to-stochastic decoders, the proposed MAC engine, and an ADC. Firstly, the decoders convert the binary input codes to stochastic codes. The input numbers of feature maps are converted to 11-bit stochastic numbers, and the weights 4-bit (both of the two have an extra sign bit). The input precision is selected based on the network which our MAC engine is applied to. According to the description of deterministic coding in Section2.1, both of them are extended to 4×11=44 bits. Then, multiplications of the stochastic numbers are completed by the AND gate array. After that, these products are converted into analog MAC result by the analog accumulate engine (ACE). Finally, the analog output is fed to a flash ADC to be converted to binary numbers to prepare for next cycle's calculation.

The ACE is composed of three parts: stochastic-to-analog converter (SAC), voltage-to-time converter (VTC) and Fig.

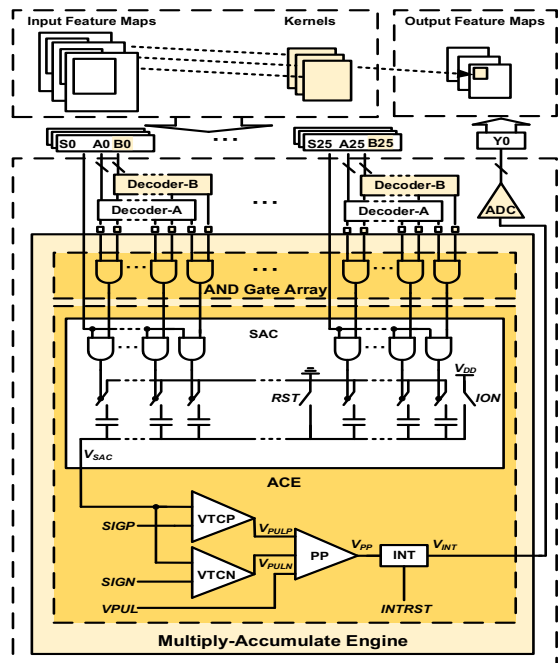

Fig. 2. Operation diagram of convolution calculation with proposed mixed-signal MAC engine.

integrating circuit (INT). SAC calculates the convolution sum of one input feature map and one kernel, VTC converts the sum into pulse duration so that sums from different feature maps can be accumulated by INT. In this work, we take 5×5 kernel (with a bias), 6 feature maps as example to explain the working process of our design. According to the sign of the input numbers Ai and Bi  (i={0,1,…,25}), SAC converts the sum of positive numbers and negative numbers to analog voltage $V_{SACP}$ and $V_{SACN}$ separately, and then VTCP and VTCN are employed to convert $V_{SACP}$ and $V_{SACN}$ into pulse signals respectively during two conversion stages, the width of which is proportional to $V_{SAC}$ (including $V_{SACN}$ and $V_{SACP}$). To calculate the difference between $V_{SACP}$ and $V_{SACN}$, we employ a pulse processing (PP) circuit with a reference pulse input $V_{PUL}$. The pulse width of $V_{PUL}$ represents the situation when input is 0. The pulse width of $V_{PULN}$ is proportional to the absolute value of negative input number, and the pulse width of $V_{PULP}$ represents the value of positive input number. To calculate $V_{SACP} - V_{SACN}$, PP circuit does subtraction between $V_{PUL}$ and $V_{PULN}$ and then adds the result with $V_{PULP}$ and obtains a signed addition result.

Finally, $V_{PP}$ controls the charging time of the INT so that the output voltage is proportional to the pulse width of $V_{PP}$. As data of different feature maps continuously run to MAC engine one after another, each convolution sum is added up to the output of INT, which performs the accumulation operations. The sequence diagram of proposed mixed-signal MAC engine is showed in Fig. 3.

Charge redistribution (CR) technique is applied to multiply-accumulate (MAC) circuit [10] [11] to improve energy efficiency. Thus an SAC (Fig. 4) based on CR is proposed in our MAC engine to do addition.

The SAC consists of a capacitor array switched by the output of AND gate array, and all capacitors are of the same size. To distinguish positive inputs from negative ones, the operation of SAC is composed of two stages: POS and NEG. Both of them can be divided into three phases: RESET, CHARGE and SHARE. In the RESET phase, *RST* is set to be valid, both the top and bottom plate of capacitors are connected to *GND* and all capacitors are discharged. In the CHARGE phase, the switches controlled by negative DIN are off, and positive DIN set the other switches off or on according to the value of Di[j]. After that, *ION* is set to be valid so that top plates of all capacitors are connected to $V_{DD}$, the capacitors with bottom plate connected to *GND* are charged. In the SHARE phase, *PUP* is set to be high, all bottom plates are connected to *GND* and all top plates are connected with each other, charge is shared over all capacitors. Then the output voltage of $V_{SAC}$ is proportional to the sum of positive DINs. The operations in RESET phase and SHARE phase of NEG stage are same as POS stage. The only difference happens in CHARGE phase, the capacitors with negative DINs are charged so that $V_{SAC}$ is proportional to the sum of negative DINs' absolute value. Thus we get the sum of positive numbers ($V_{SACP}$) and negative numbers ($V_{SACN}$) separately after the two stages.

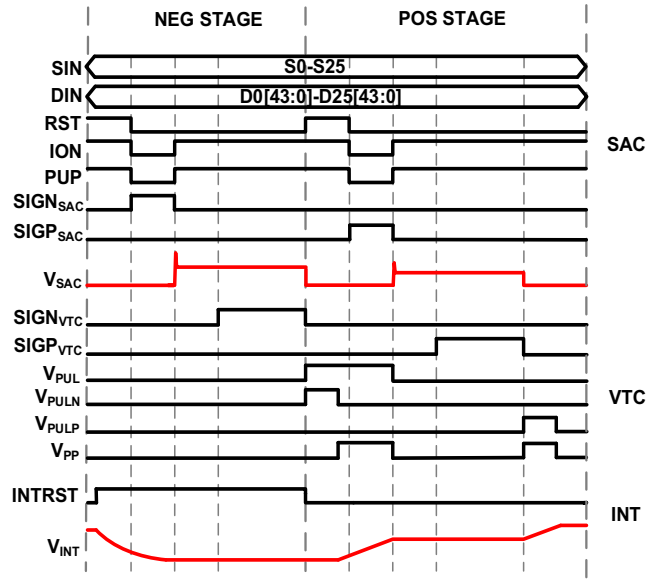

Fig. 3. Sequence diagram of mixed-signal MAC engine.

The output of SAC need to be accumulated using an integrating circuit proposed in the next part. Thus we employ VTC to convert the analog voltage into pulse signals. There are various kinds of VTC architectures with different characteristics. The design based on basic current starved inverter [12] shows low power consumption but poor linearity. Another kind of VTC using comparator and slope generator [13] has the feature of high precision but is also more complicated and limited to low operation frequency. The design of [14], which makes a tradeoff between energy efficiency and linearity, is therefore applied to our design, as shown in Fig. 5(a)(b). VTC works in two phases to implement the conversion. During sampling phase, $EN$=1, $ENB$=0, $I_1$=0, $I_0$=$I_{CNS}$ ($I_{CNS}$ is a constant current here), the capacitor is charged to $V_{IN}$. Then comes the discharging phase, $EN$=0, $ENB$=1, $I_0$=0, $I_1$=$I_{CNS}$, the discharging current $I_1$ is equal to $I_0$ when the frequency of $EN$ is high enough. A constant current leads to manageable discharging time, so that the output pulse

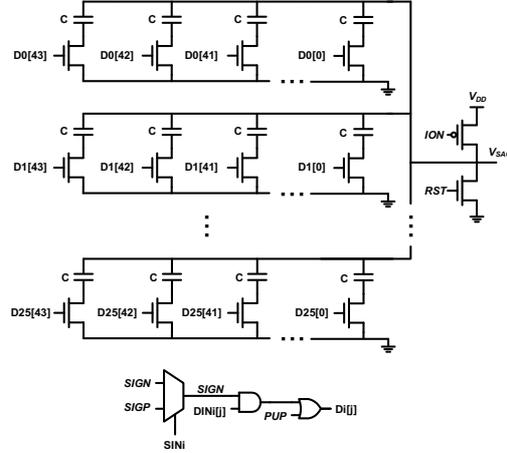

Fig. 4. Circuit of stochastic-to-analog converter (SAC).

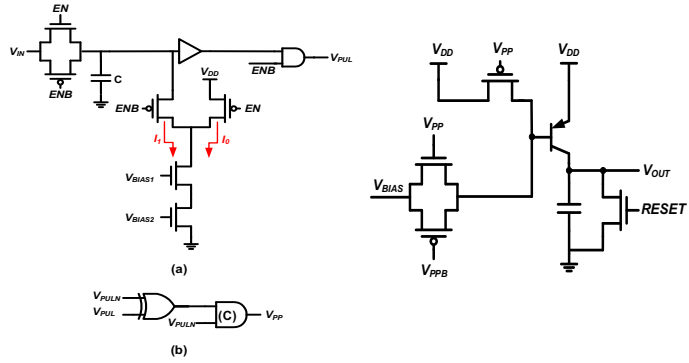

Fig. 5. (a) Circuit of voltage-to-time converter (VTC); (b) Circuit of pulse processing (PP) module (c) Integrating circuit (INT).

width is proportional to the input voltage $V_{IN}$.

To calculate the difference between $V_{SACP}$ and $V_{SACN}$, we employ a PP circuit with a reference pulse input $V_{PUL}$. PP circuit does subtraction between $V_{PUL}$ and $V_{PULN}$ with an XNOR gate and then adds the result with $V_{PULP}$ using an OR gate. The output of PP circuit is then fed to the INT.

Fig. 5(c) describes the integrating circuit in this work. $V_{PP}$ is the output of VTC part, which is a pulse signal whose width is proportional to $V_{SAC}$. The capacitor is charged when $V_{PP}$ is valid. When $V_{PP}$ turns to low, bipolar will be cut off and charging stops. The convolution sum of 6 feature maps can be added together sequentially through this part. This structure consumes less power within an acceptable precision compared to conventional integrator based on amplifier.

## 4. Simulation Results

The proposed mixed-signal MAC engine is designed under 28nm CMOS technology and operates at 1V supply voltage with 25MHz clock, and the overall power consumption is 20.12μW.

To guarantee the linearity of VTC, the output voltage of SAC should be above 0.35V. Fig. 6(a) shows the simulation result of SAC, which demonstrates the transfer characteristic between output voltage and input number, the output range is from 0.41V to 1.0V. Fig.6(b) displays the transfer characteristic between output pulse width and input voltage of VTC. Good linearity is shown when the input voltage is between 0.35V and 1.0V, which perfectly meets the requirements of system. The output curve of integrating circuit is shown in Fig. 6(c), the charge current is always stable in the structure so that the output linearly increases with the input voltage. Fig. 6(d) shows the transfer characteristic and nonlinearity of MAC engine, the absolute value of maximum error is below 0.2mV, which is much smaller than quantization step length.

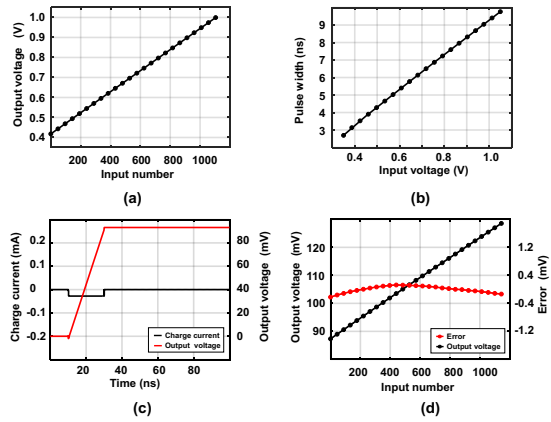

Fig. 6. Simulation result. (a) Transfer characteristic of SAC; (b) Transfer characteristic of VTC; (c) Transfer characteristic of INT; (d) Transfer characteristic of MAC engine and error in VINT versus input number

Overall, the MAC engine consumes 5.03 pJ per 26-input MAC operation. The energy efficiency of the proposed MAC engine achieves 10.14 TOPS/W.

## 5. Summary

A 44-bit, 26-input, mixed-signal SC-based MAC engine has been proposed in this paper, which can reduce the power consumption of convolution calculation. To reduce the latency, parallel SC structure is employed. Then we address the high-fan-in addition issue of parallel SC MAC engine by processing addition and accumulation in analog domain. Moreover, convolutional sum from different feature maps can be accumulated by INT, so that we can reduce the times of accessing to the external storage. The simulated energy per MAC operation is 5.03 pJ, which is beneficial to low power applications.

**Acknowledgments**

This work was supported by National Natural Science Foundation of China (Grant No.61834001 and No.61421005) and the 111 Project (B18001).